\documentclass[
   ,final            
  ]
  {aipproc}

\layoutstyle{8x11single}

\begin{document}

\title{Forward Jets and Forward-Central Jets at CMS}


\author{Niladri Sen, on behalf of the CMS Collaboration}{
  address={Deutsches Elektronen-Synchrotron, Notkestra\ss e 85, Hamburg, Germany}
}

\begin{abstract}
We report on cross section measurements for inclusive forward jet production and for the simultaneous production of a forward and a central jet in $\sqrt{s} = $7 TeV $pp$-collisions. Data collected in 2010, corresponding to an integrated luminosity of 3.14 pb$^{-1}$, is used for the measurements. Jets in the transverse momentum range p$_{T}$ = 35 - 140 GeV/c are reconstructed with the anti-k$_{T}$~(R = 0.5) algorithm. The extended coverage of large pseudo-rapidities is provided by the Hadronic Forward calorimeter (3.2 $< | \eta | <$ 4.7), while central jets are limited to $|\eta| <$ 2.8, covered by the main detector components. The two differential cross sections are presented as a function of the jet transverse momentum. Comparisons to next-to-leading order perturbative calculations, and predictions from event generators based on different parton showering mechanisms ({\sc pythia} and {\sc herwig}) and parton dynamics ({\sc cascade}) are shown. 
\end{abstract}

\maketitle


\section{Introduction}

Jet production in hadron-hadron collisions is sensitive to underlying partonic processes, initial- and final-state radiation ({\sc isr} and {\sc fsr}), and to the parton density functions ({\sc pdf}s) of the colliding hadrons. Measurements of jet cross sections at previous colliders are well described over several orders of magnitude by perturbative calculations~\cite{bib:cdf}~\cite{bib:d0}. However, the jets were limited to central pseudo-rapidities ($|\eta| < $2.4), where the momentum fraction of the incoming partons ($x_{1}$,~$x_{2}$) were of the same order. Jets at large pseudo-rapidities (i.e., forward or backward jets, $|\eta| > $3 ) result from interactions between colliding partons with differing momentum fractions (e.g. $x_{1} << x_{2}$), allowing us to investigate QCD-effects at small-$x$. At such small-$x$ values, {\sc pdf}s are well constrained by {\sc dis} data, but there might be additional effects that play a role. We expect signals of parton dynamics beyond the standard {\sc dglap} evolution (e.g., {\sc BFKL} or {\sc CCFM}), and saturation effects are foreseen. Moreover, forward jets are of interest in vector-boson-fusion processes, which is one of the mechanisms for Higgs boson production.

The Large Hadron Collidor ({\sc lhc}) is a proton-proton collider with a beam energy of 7 TeV and a design luminosity of $L = $34 cm$^{-2}$s$^{-1}$, designed to explore a new energy scale. At the collider, momentum fraction of the proton ($x$) carried by the partons can become very small and the parton densities become large.  Additionally, the probability of more than one partonic interaction per event increases for collision energies produced at the {\sc lhc}, and probing the forward region opens up opens up a large phase space for {\sc qcd} emissions. Thus, the {\sc lhc} provides the perfect opportunity to study small-$x$ physics and {\sc qcd} effects through measurements of forward and forward-central jet production. Here, a measurement of the cross section of inclusive forward jets~\cite{bib:pas003} and of the production of a forward jet in conjunction with a central jet~\cite{bib:pas006}, using data collected by the Compact Muon Solenoid ({\sc cms}), in $\sqrt{s} =$ 7 TeV $pp$-collisions at the {\sc lhc} is presented.

\section{The CMS Detector}
The CMS experiment~\cite{bib:jinst}, located on the French-side of CERN, operates a multi-puprpse detector to study $pp$ collisions at the LHC. The detector covers a solid angle approaching 4$\pi$ and incorporates vertex, calorimeter and muon chambers. The tracking system covers the pseudo-rapidity range -2.5 $< \eta < $ 2.5 and the calorimeter system covers the range \\ - 5 $< \eta < $ 5. The Hadronic Forward Calorimeter ({\sc HF}), made of iron absorbers embedded with radiation hard quartz fibres, is used for measuring forward jets and energy, providing an almost hermetic coverage upto $|\eta| \sim$ 5. 

\section{Cross Section Measurement For Inclusive Forward Jet Production}
Jets are built from calorimeter information using the infrared and collinear safe anti-k$_{T}$ (R=0.5) jet clustering algorithm~\cite{bib:antiKt}. A single jet trigger (p$_{T} >$ 15 GeV), fully efficient in region of the measurement (p$_{T} >$ 35 GeV) selects jets from the data sample. Jet quality criteria are applied to remove unphysical energy deposits, and the selection requirements ensure that the jets are well contained within the fiducial acceptance, i.e. their reconstructed axis is within 3.2 $<  |\eta| <$ 4.7~\cite{bib:jme003}. 

The jet cross section is fully corrected for detector reconstruction effects via a bin-by-bin correction factor calculated from simulated samples. The number of jets, N$_{jets}$, are binned in transverse momentum (p$_{T}$) and pseudo-rapidity ($\eta$). The final differential inclusive jet cross-section is:

\begin{equation}
\frac{d^{2}\sigma}{dp_{T}d\eta} = \frac{C_{unfold}}{L} \cdot \frac{N_{jets}}{\Delta p_{T} \cdot \Delta{\eta}}
\end{equation}

where C$_{unfold}$ is the correction factor accounting for detector effects (e.g. migrations, resolution), and $\Delta p_{T}$ and  $\Delta{\eta}$ are the bin widths in $p_{T}$ and $\eta$ respectively. 

Both experimental and theoretical sources of uncertainties have been considered. The dominant experimental systematic uncertainty is the accuracy of the jet energy scale ({\sc jes}), which is 20\%--30\% in all p$_{T}$ bins of the measured cross section. Additional sources include the p$_{T}$ resolution for bin-by-bin corrections (3\%--6\%) and the luminosity measurement (4\%). The theoretical uncertainties are estimated by checking non-perturbative effects through {\sc pythia} and {\sc herwig} comparisons , ascertaining the impact of different {\sc pdf}s and the variation of $\mu_{r}$ and $\mu{f}$ by a factor of 2.

\begin{figure}
\label{jetcross}
\rotatebox{90}{\includegraphics[width=0.4\textwidth]{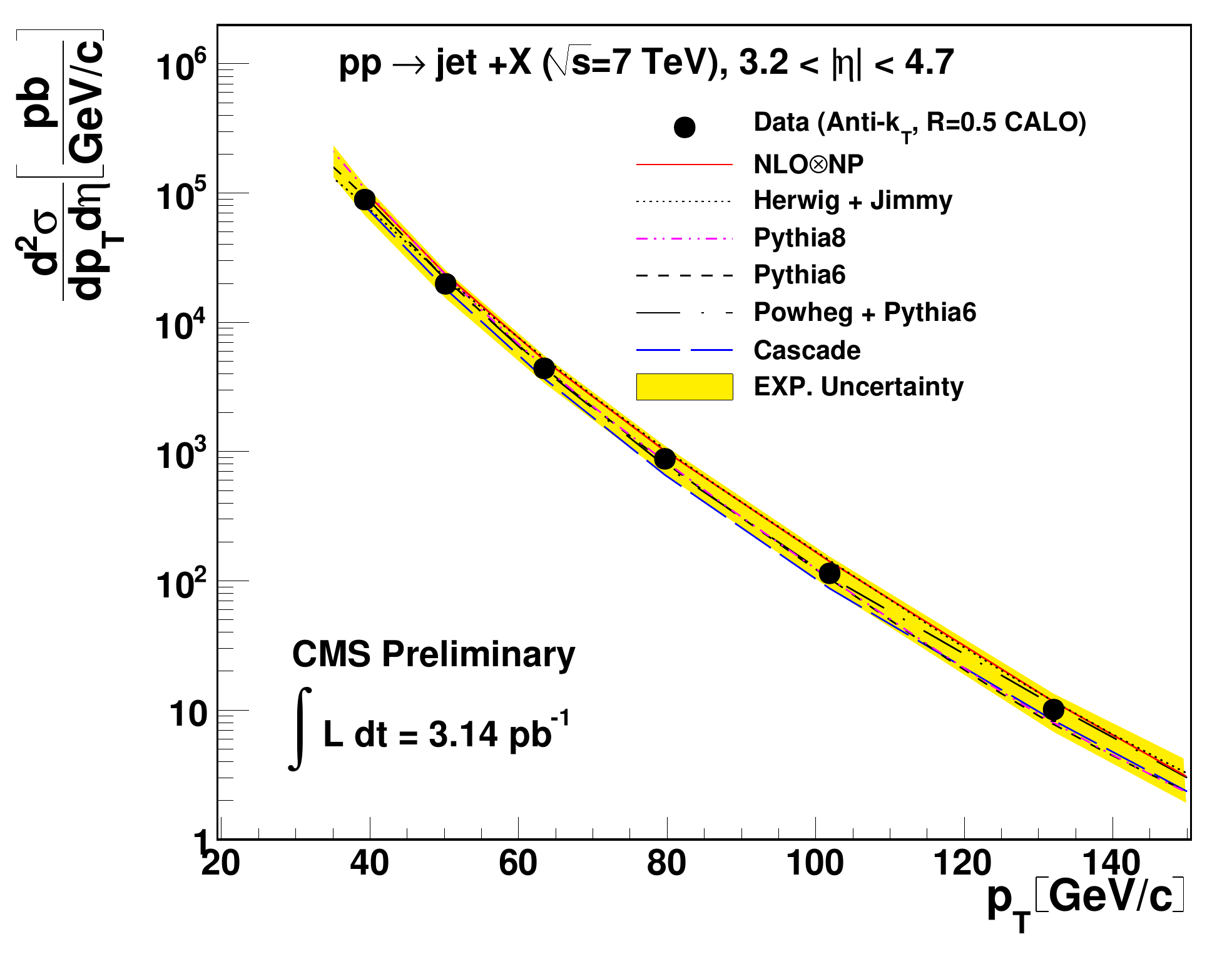}}
\caption{Inclusive jet cross section measured at forward pseudo-rapidities (3.2 $ < |\eta| < $ 4.7), fully corrected and unfolded, compared to various hadron-level predictions. The error band represents the experimental systematic uncertainty.}
\end{figure}

\begin{figure}
\label{jetfraction}
\rotatebox{90}{\includegraphics[width=0.4\textwidth]{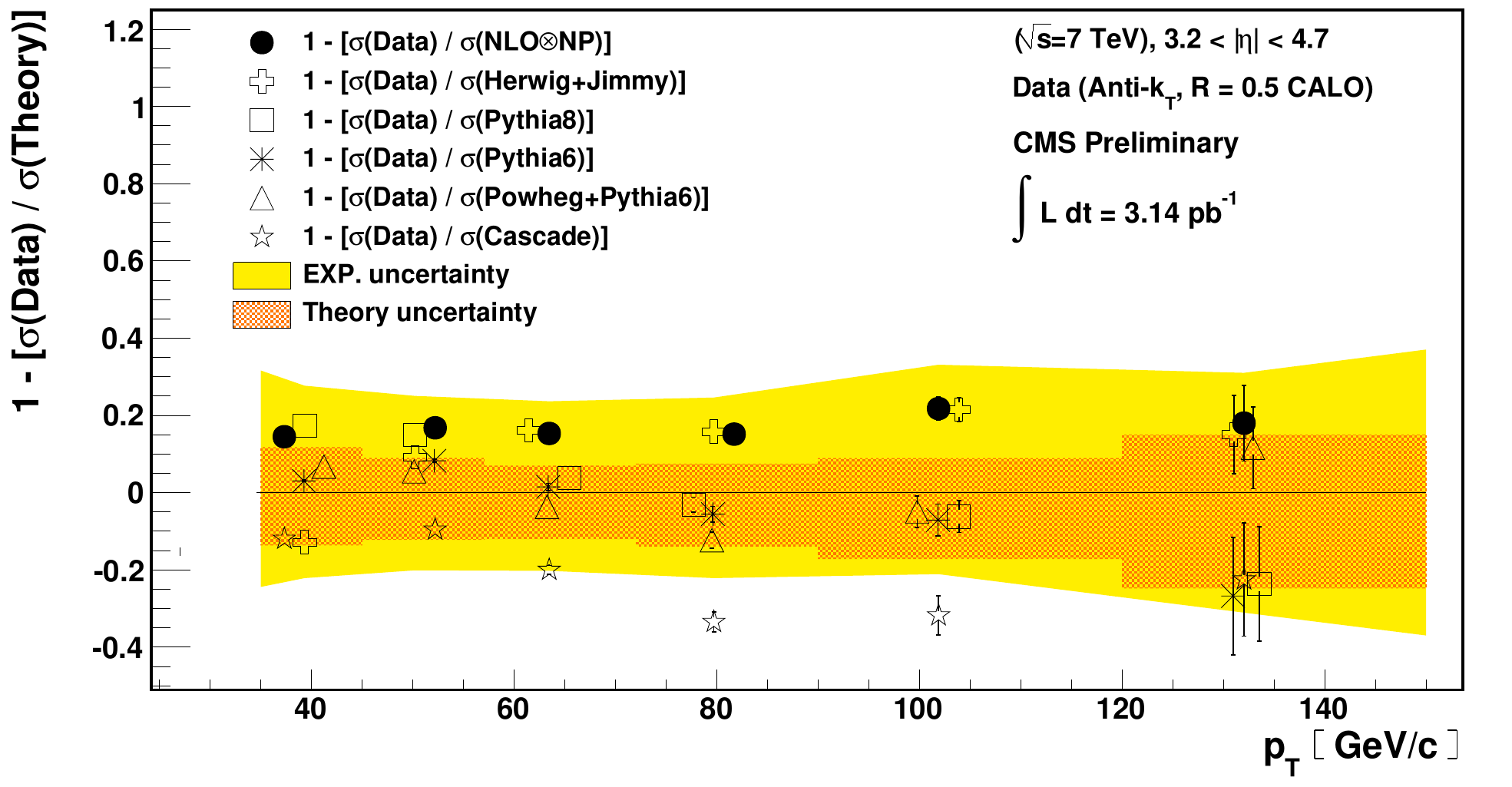}}
\caption{Fractional differences between forward jet spectra and theoretical predictions. The error-bars on the data points show statistical uncertainties. The error bands represent the systematical and theoretical uncertainties.}
\end{figure}

Figure~\ref{jetcross} shows the fully corrected jet cross section as a function of p$_{T}$, in comparison to various theoretical models. The yellow-band indicates the experimental uncertainty. Figure~\ref{jetfraction} shows the fractional difference between the experimental jet cross-section and the theoretical predictions. Within the uncertainties, the  hadron-level predictions are in good agreement with the data. The exception is  {\sc cascade}, where shape variations at high p$_{T}$ are observed between the measured forward jet spectra and the prediction. 

\section{Cross Section Measurement for Simultaneous Production of a Forward and a Central Jet}
The jet clustering algorithm, widths in the p$_{T}$ spectrum, and selection criteria are the same in this measurement as for the inclusive jet cross section analysis. The only difference is the additional requirement of a well-reconstructed central ($|\eta| <$ 2.8) jet . That is, an event is accepted if there are at least two reconstructed jets with p$_{T} > $ 35 GeV, one with its axis within the central region ($|\eta| <$ 2.8), and the other within the {\sc hf} (3.2 $ < |\eta| < $ 4.7). If there is more than one jet present in either of the regions, the leading jet is considered.

The jet cross-sections are obtained using a bin-by-bin correction as for the inclusive jet production analysis. In this case, however, the simulated samples were re-weighted at hadron level to describe the measured data distributions. The {\sc jes} remains as the dominant systematic uncertainty ($\sim$ 25\%). The model dependence is the second largest uncertainty, estimated by the variation of bin-by-bin correction factors from different {\sc mc}s (5\% -- 15\%).

\begin{figure}
\label{central}
\includegraphics[width=0.4\textwidth]{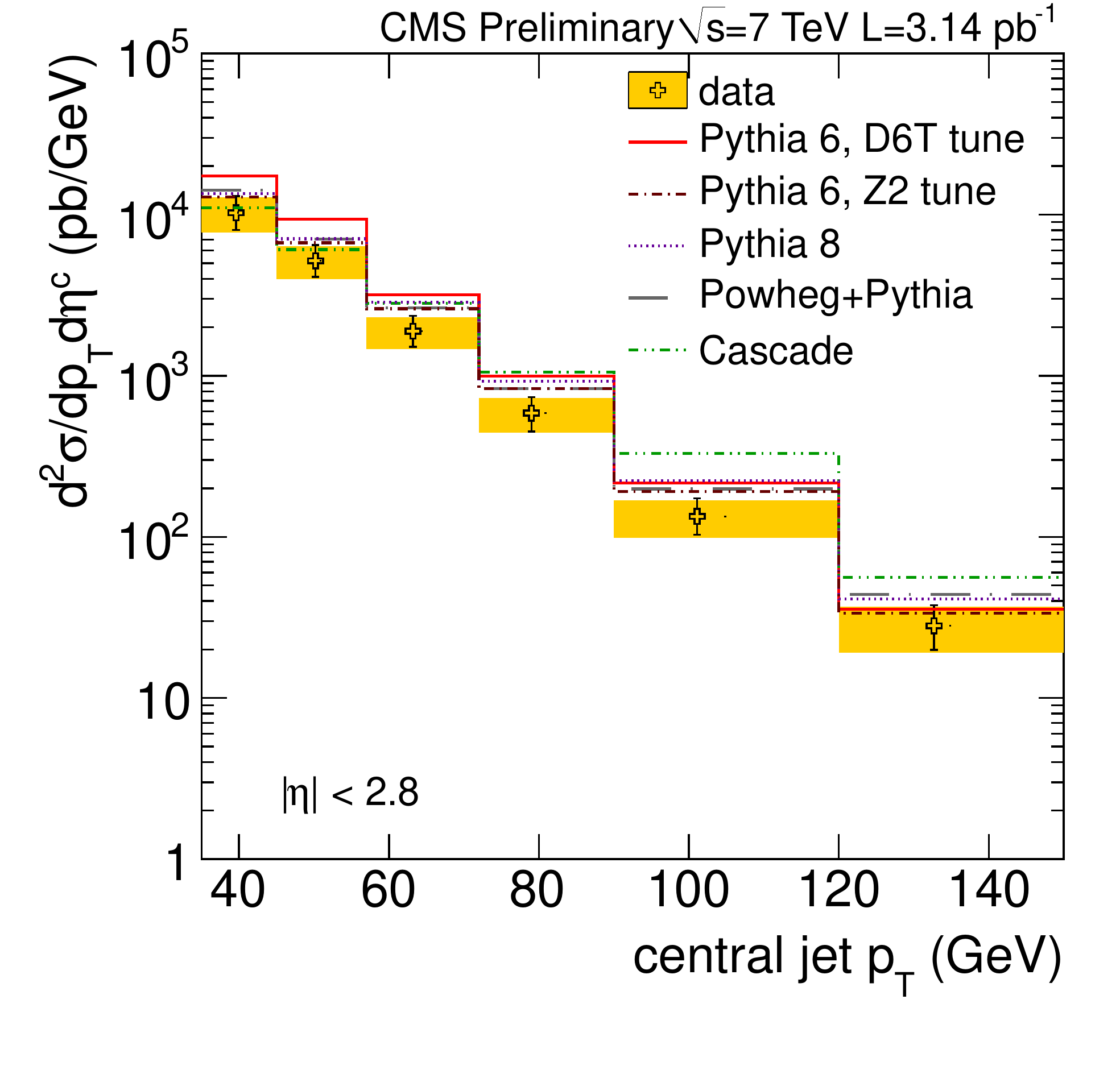}
\includegraphics[width=0.4\textwidth]{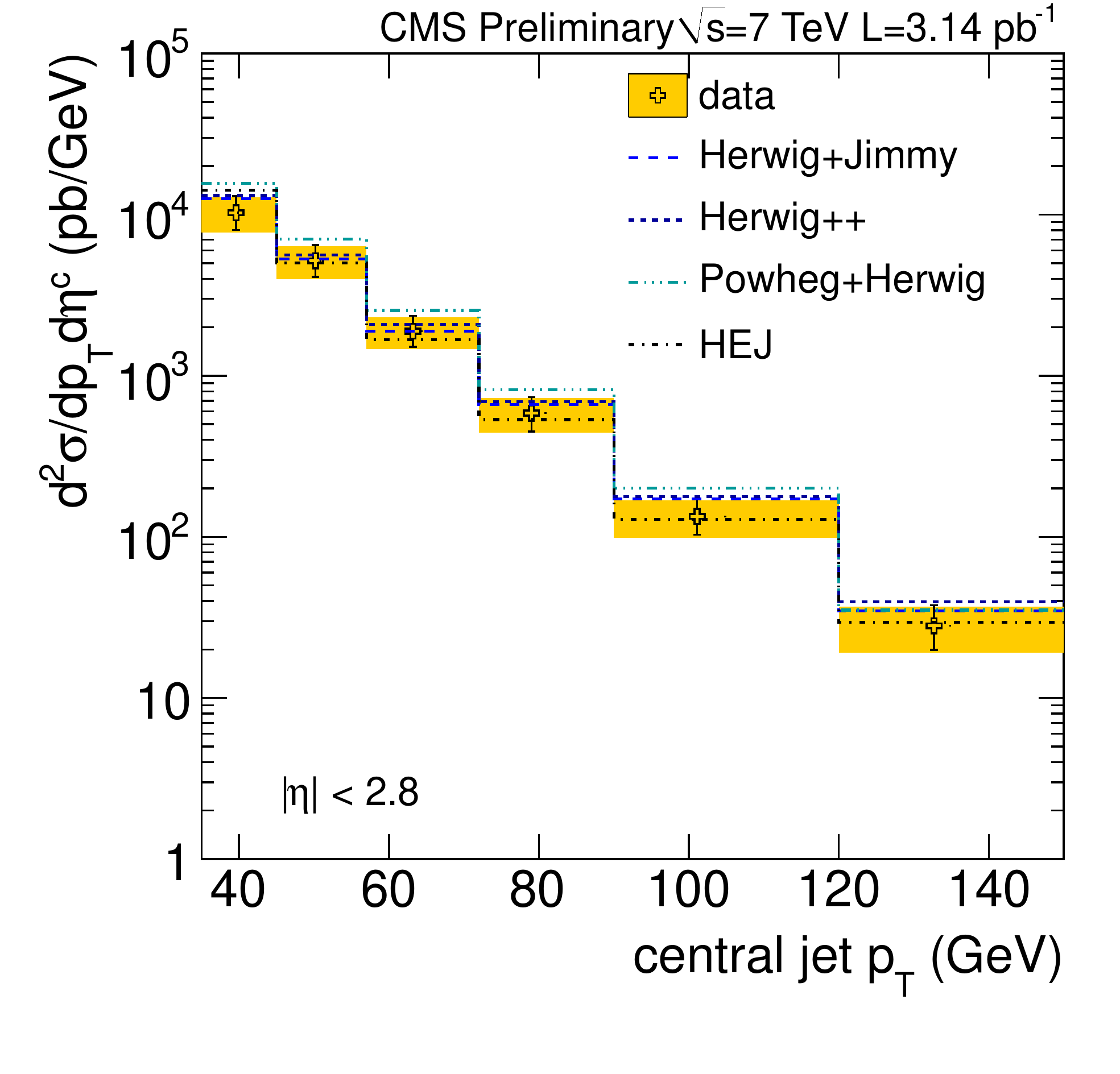}
\caption{Fully corrected p$_{T}$-differential jet cross-section for the central region ($|\eta| <$ 2.8) compared to various event generators, {\sc pythia} and {\sc cascade} (left), {\sc herwig} and {\sc hej} (right). The error-bars on the data points show statistical uncertainties. The error bands represent the systematical.}
\end{figure}

\begin{figure}
\label{forward}
\includegraphics[width=0.4\textwidth]{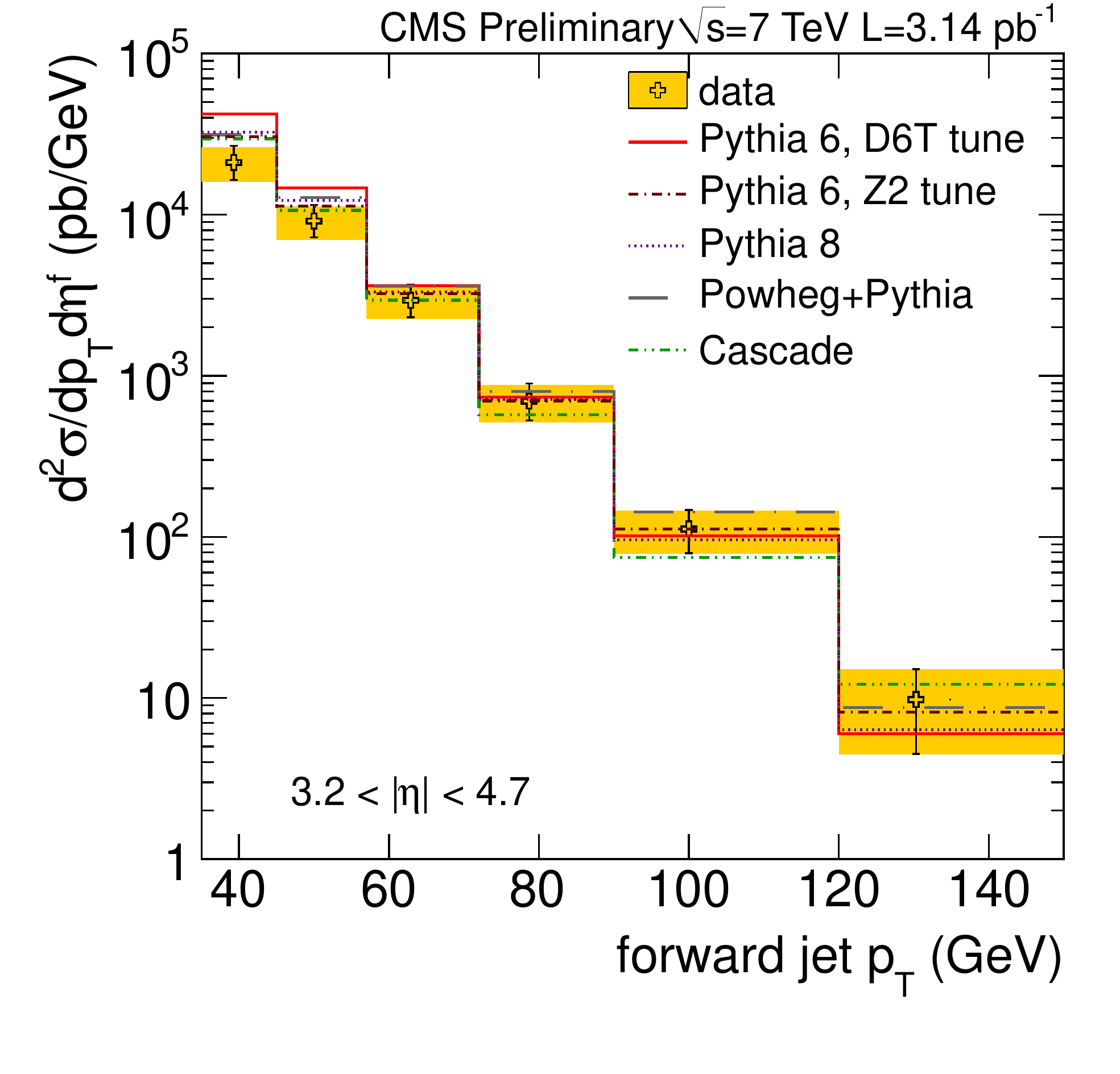}
\includegraphics[width=0.4\textwidth]{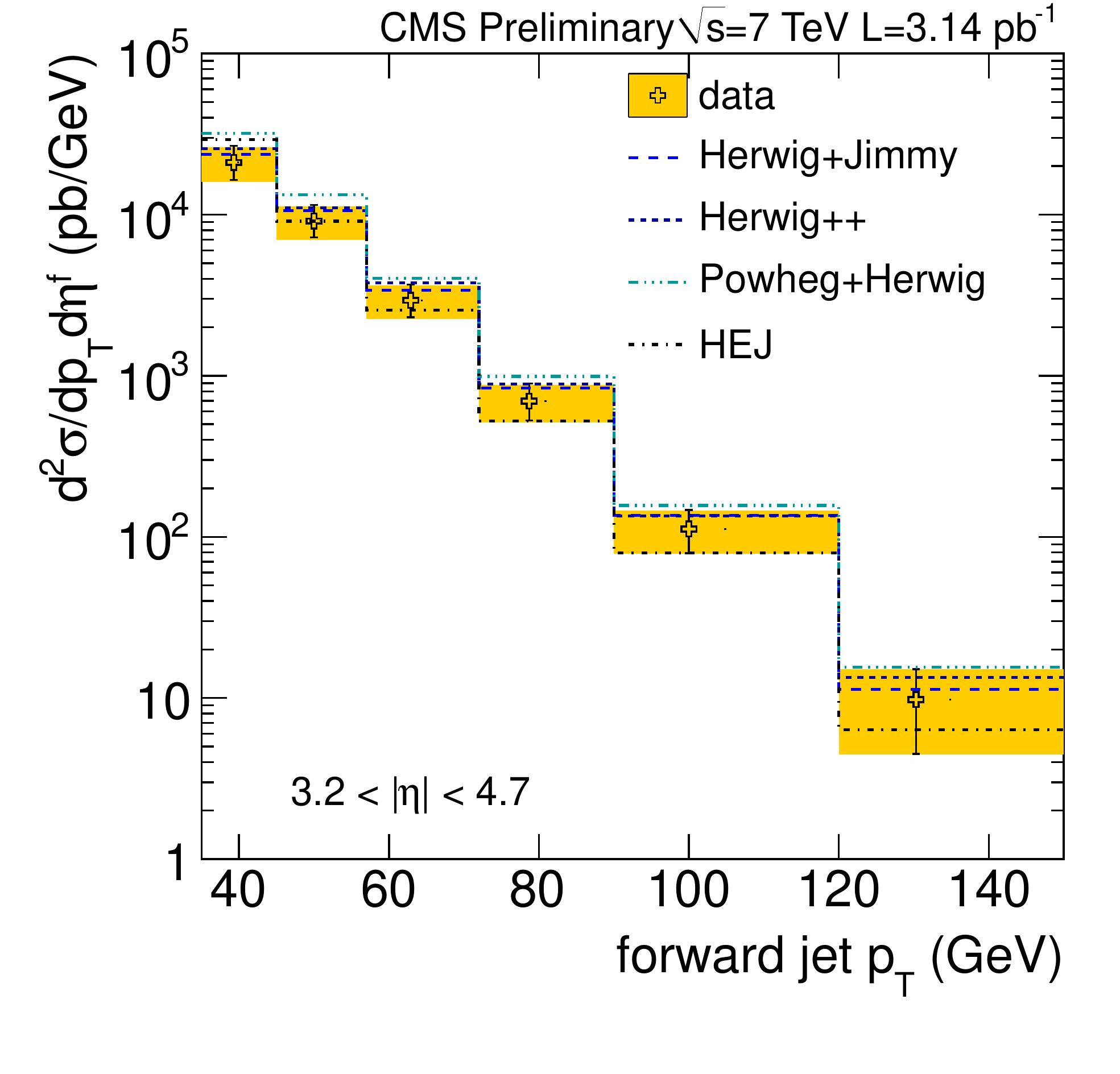}
\caption{Fully corrected p$_{T}$-differential jet cross-section for the forward region (3.2 $ < |\eta| < $ 4.7) compared to various event generators, {\sc pythia} and {\sc cascade} (left), {\sc herwig} and {\sc hej} (right). The error-bars on the data points show statistical uncertainties. The error bands represent the systematical uncertainties.}
\end{figure}

The fully corrected cross section for simultaneous production of at least one central and one forward jet is measured as a function of jet p$_{T}$. Figures ~\ref{central}--~\ref{forward} present the measurement with the corresponding hadron-level predictions in contrast, for central jets and forward jets respectively. The yellow bands indicate the experimental systematic uncertainties summed in quadrature and the error bars represent the statistical uncertainty. The plots on the left show some discrepancies between {\sc pythia} and data, especially in the central region. {\sc herwig}, which uses angular-ordered parton showers, describes the data better, for both regions of pseudo-rapidity. 


\section{Conclusion}
We present a measurement of jet production in the forward pseudo-rapidity range 3.2 $ < |\eta| < $ 4.7, and the cross section for the simultaneous production of one central and one forward jet, using 3.14 pb$^{-1}$ of {\sc cms} data collected during the early $\sqrt{s} =$ 7 TeV pp-collisions. Within the current experimental and theoretical uncertainties, perturbative calculations reproduce the measured inclusive forward jet cross section well. The data-model comparison of the forward-central jet measurement show that only some calculations are in reasonable agreement with data. Both measurements are a first test of perturbative {\sc qcd} calculations in the forward region, providing the basis for further investigation of this interesting region of phase space. 

\section{Acknowledgements}
Copyright CERN for the benefit of the CMS Collaboration.




\bibliographystyle{aipproc}   

\bibliography{panic_11_proceedings}

\hyphenation{Post-Script Sprin-ger}
\begin{thebibliography}{7}
\expandafter\ifx\csname natexlab\endcsname\relax\def\natexlab#1{#1}\fi
\providecommand{\enquote}[1]{``#1''}
\expandafter\ifx\csname url\endcsname\relax
  \def\url#1{\texttt{#1}}\fi
\expandafter\ifx\csname urlprefix\endcsname\relax\def\urlprefix{URL }\fi
\providecommand{\eprint}[2][]{\url{#2}}

\bibitem[Collaboration(2008{\natexlab{a}})]{bib:cdf}
CDF Collaboration, \emph{Phys. Rev.} \textbf{D78} (2008{\natexlab{a}}).

\bibitem[Collaboration(2008{\natexlab{b}})]{bib:d0}
D0 Collaboration, \emph{Phys. Rev. Lett.} \textbf{101} (2008{\natexlab{b}}).

\bibitem[Collaboration(2011{\natexlab{a}})]{bib:pas003}
CMS Collaboration, Measurement of forward jets in proton-proton collisions at 7
  TeV (2011{\natexlab{a}}), CMS Physics Analysis Summary FWD-10-003.

\bibitem[Collaboration(2011{\natexlab{b}})]{bib:pas006}
CMS Collaboration, Cross section measurement for simultaneous production of a
  central and a forward jet in proton-proton collisions at 7~TeV
  (2011{\natexlab{b}}), CMS Physics Analysis Summary FWD-10-006.

\bibitem[Collaboration(2008{\natexlab{c}})]{bib:jinst}
CMS Collaboration, \emph{The CMS experiment at the CERN LHC}, JINST 3:S08004,
  2008{\natexlab{c}}.

\bibitem[M.~Cacciari(2008)]{bib:antiKt}
G.Soyez, M.~Cacciari, G.P.~Salam, \emph{JHEP} \textbf{04} (2008).

\bibitem[Collaboration(2007)]{bib:jme003}
CMS Collaboration, Performance of jet algorithms in CMS (2007), CMS Physics
  Analysis Summary JME-07-003.

\end{thebibliography}

\IfFileExists{\jobname.bbl}{}
 {\typeout{}
  \typeout{******************************************}
  \typeout{** Please run "bibtex \jobname" to optain}
  \typeout{** the bibliography and then re-run LaTeX}
  \typeout{** twice to fix the references!}
  \typeout{******************************************}
  \typeout{}
 }

\end{document}